\documentclass[times, 10pt, twocolumn]{article} 
\usepackage{latex8}
\usepackage{times}
\usepackage{amsmath,amssymb,amsfonts,amsthm,latexsym,eucal}
\usepackage{url,graphicx,amssymb,amsfonts}
\usepackage{fullpage}
\usepackage{hyperref}

\newtheorem{definition}{Definition}
\newtheorem{theorem}{Theorem}

\newcommand{\Exp}{\mathbb{E}}

\newcommand{\bigdisjointunion}{\dot{\bigcup}}

\newcommand{\OPT}{\textrm{OPT}}

\newcommand{\cw}{\mathbf{cw}}

\newcommand{\remove}[1]{}

\title{Randomized Work-Competitive Scheduling for Cooperative Computing on $k$-partite Task Graphs} 
\author{
Chadi Kari\\
	\textsf{chadi@engr.uconn.edu} \and
Alexander Russell\\
  \textsf{acr@cse.uconn.edu} \and
  Narasimha Shashidhar\\
\textsf{karpoor@cse.uconn.edu} \and
Department of Computer Science and Engineering\\
University of Connecticut, Storrs, CT}

\date{}

\begin{document}

\maketitle

\section{Introduction}
A fundamental problem in distributed computing is the problem of cooperatively executing a given set of tasks in a dynamic setting. The challenge is to minimize the total \emph{work} done and to maintain efficiency in the face of dynamically changing processor connectivity. In this setting, \emph{work} is defined as the total number of tasks performed (counting multiplicities) by all the processors during the course of the computation.

In this scenario, we are given a set of $t$ tasks that must be completed in a distributed setting by a set of $p$ processors where the communication medium is subject to failures. We assume that the $t$ tasks are similar, in that they require the same number of computation steps to finish execution. We further assume that the tasks are idempotent - executing a task multiple times has the same effect as a single execution of the task. The tasks have a dependency relationship defined among them captured by a task dependency graph. 

The dynamics of the communication medium determine a processor's ability to communicate with other processors. Effectively, this partitions the processors into groups. Processors that can communicate with each other are said to belong to the same group. No communication is possible between processors in different groups. Each processor of a group is aware of all the tasks completed by the members of the group. The dynamic changes in the communication medium leads to a reconfiguration, i.e. a new partition of processors into groups. This new group of processors share knowledge of all the tasks that have been completed among them so far and then proceed to continue executing the remaining tasks from their pool of incomplete tasks until the next reconfiguration. \\

This processor group reconfiguration and task execution may be treated as if they were determined by an adversary. Thus, the adversary in our model performs two basic operations: reconfigures the processors into groups and also allocates the work quota for each group of processors before the next reconfiguration. The work quota is the number of tasks that can be completed by the group before the next reconfiguration takes place. While the adversary controls the \emph{number} of tasks that a group can perform, he does not dictate \emph{which} tasks (the identity of the tasks) the group can perform. \\ \indent In this setting, the tasks have dependencies defined among them captured by a directed acyclic task graph ($t$-DAG) which is a $k$-partite task graph. Given a group of processors and the tasks known to be completed by them, an algorithm in this setting decides on the next incomplete task to be completed by this group. Each processor continues to execute tasks from the given set of $t$ tasks until it is aware that all tasks have been completed or runs out of it's allocated work limit. Hence, given $p$ processors and $t$ tasks, any algorithm must execute at least $\Omega\left(t \cdot p\right)$ tasks in the scenario where all the processors are disconnected for the entire computation while any reasonable algorithm would only incur $O(t)$ \emph{work} in the completely connected case. Hence, we treat this problem in an on-line setting and pursue competitive analysis where the performance of our algorithm is compared against that of the omniscient offline algorithm which has complete knowledge of all the future changes to the communication medium. Our setting is a generalization of the problem in ~\cite{GRS_STOC03,GRS_SIAM05} since the tasks are no longer independent but have dependencies defined among them. We show that for this setting more pessimistic bounds hold.
%
%
%
\section{Our Results}
Georgiou, Russell, and Shvartsman~\cite{GRS_SIAM05} performed competitive analysis and showed a simple randomized scheduling algorithm $RS$ (Random Select) whose competitive ratio is tight. Their work also introduced a notion of \emph{computation width}, which associates a natural number with a history of changes in the communication medium, and shows both upper and lower bounds on competitiveness in terms of this quantity. Specifically, they showed that their simple randomized scheduling algorithm obtains the competitive ratio $\left(1+\cw/e\right)$, where $\cw$ is the computation width of the computation pattern determined by the dynamics of the communication medium. 
We follow on the work done in ~\cite{GRS_SIAM05}. We study a natural generalization of the problem where the tasks to be completed are not independent of each other but have a $k$-partite dependency relationship defined among them. Each partition of the vertices (tasks) of the $k$-partite task graph is said to belong to a level. Independent tasks belong to the first level, tasks dependent on the first level tasks are at the second level and so on. The $k$-partite task graphs that we consider in our problem are a special kind of task graphs where every task at level $l_{i+1}$ is dependent on every task at level $l_i,~i=1,\ldots,k-1$ (i.e, complete set of directed edges from level $l_i$ to level $l_{i+1},~i=1,\ldots,k-1$). We present a simple randomized algorithm for $p$ processors cooperating to perform $t$ known tasks where the dependencies between them are defined by a $k$-partite task dependency graph with processors subject to a dynamic communication medium. We pursue competitive analysis and show that pessimistic bounds hold in this case. \\
\noindent
Our algorithm Modified-$RS$ extends the algorithm \emph{Random Select}($RS$) presented in~\cite{GRS_SIAM05}. Modified-$RS$ is a simple randomized scheduling algorithm whose competitive ratio depends on the \emph{computation width}~\cite{GRS_SIAM05} and the nature of dependencies among the tasks captured by the task graph. 
We show in section ~\ref{sec:upperbound} that algorithm Modified-$RS$ is $\left(1+ cw\left(1-\alpha + \frac{\alpha}{e^{\frac{1-\alpha}{\alpha} c+1}}\right)\right)$-competitive for any computational $(p,t)$-DAG and for a 2-level task $t$-DAG  where, $cw$ is the \textbf{computation width} of the computational pattern, $\alpha \in (0,1]$ denotes the fraction of tasks in the first level $l_1$ and $c = \frac{1}{\frac{1}{e} + o(1)}$. This competitive ratio matches the lower bound we show in section ~\ref{sec:lowerbound} and therefore is tight.  We then extend our analysis to any $k$-level task $t$-DAG. We show that Modified-$RS$ is $\left(1+ cw\left((1-\alpha_1) +  \frac{\alpha_1 }{e^{\frac{\alpha_k}{\alpha_1} c^{a_k}+a_k}}\right)\right)$-competitive for any computational pattern and for any $k$-level task $t$-DAG where, $\alpha_i \in (0,1]$ and $c=\frac{1}{\frac{1}{e} + 1}$ and where $a_i, i=1 .. k$ is a sequence defined as follows, $a_1 = 1, a_{i+1} = \frac{\alpha_i}{\alpha_1}c^{a_i} + a_i$. Here, $\alpha_i \in (0,1]$ is the fraction of tasks at level $l_i,~i=1,\ldots,k$. $cw$ stands for the computation width of the computational pattern and $c_i>0$. We also show that this result is tight as it matches the lowerbound we show before.
\noindent
When all the tasks given are independent i.e. the task $t$-DAG has only one level ($\alpha=1$) the competitive ratio collapses to $\left(1+ cw/e\right)$, the bound offered by ~\cite{GRS_SIAM05}. Hence, our results subsume the results of ~\cite{GRS_SIAM05}. \\
\section{Model and Definitions}
\label{sec:Model}

The problem is defined in terms of $p$ asynchronous processors and $t$ tasks with unique identifiers, initially known to all processors. For our purposes the tasks are idempotent and similar, i.e., each task requires the same number of computation steps.

\begin{definition} 
A $t$-DAG is a directed acyclic $k$-partite graph $G=(V,E)$, where $ V = \bigdisjointunion_{l=1}^{k} V_l=[t] = \{ 1 \dots t\}$. Edge $e = (t_i^l,t_j^{l+1})\in E,~l=1,\ldots,k-1$, $i\neq j$ if and only if task $t_j^{l+1}$ depends on task $t_i^l$. We write $t_i^l < t_j^{l+1}$ if task $t_j^{l+1}$ depends on task $t_i^l$. Here, $\bigdisjointunion$ stands for disjoint union. 
\end{definition}
\noindent
We only consider task graphs where a task on level $l_{i+1}$ depends on all tasks of level $l_i$. The computation pattern i.e., the computational $(p,t)$-DAG defined below captures the behavior of the adversary that determines both the partitioning and the number of tasks allocated to each group of the partition.

\begin{definition}
A computational \textbf{$(p,t)$-DAG} is a directed acyclic graph $C=(V,E)$ augmented with a weight function $h:V\rightarrow [t]\cup \{0\}$ and a labeling $g:V\rightarrow2^{[p]}~ \backslash{}~  \{ \emptyset{} \}$ so that: 1)  For any maximal path $P=\left(v_1,v_2,\ldots,v_k\right)$ in $C$, $\sum_{i=1}^k h(v_i) \geq t$. (This guarantees that any algorithm terminates during the computation described by the DAG.) \\2) g possesses the following ``initial conditions'': $[p] = \bigcup_{v:~ in(v)=0}^{.} g(v).$\\ 3) g respects the following ``conservation law'': There is a function $\phi : E \rightarrow 2^{[p]}\backslash \{\emptyset\}$ so that for each $v \in V$ with $in(v) > 0$, $g(v) = \bigcup_{(u,v) \in E}^{.} \phi((u,v))$, and for each $v \in V$ with $out(v) > 0$, $g(v) = \bigcup_{(v,u) \in E}^{.} \phi((v,u)).$

\end{definition}

In the above definition, $in(v)$ and $out(v)$ denote the in-degree and out-degree of $v$ respectively. Finally, for the two vertices $u, v \in V$, we write $u \leq v$ if there is a directed path from $u$ to $v$; we then write $u < v$ if $u \leq v$ and $u$ and $v$ are distinct.

\begin{definition}
Given a computational DAG $C = (V,E)$ and a vertex $v \in V$, we define the \textbf{predecessor graph} at $v$, denoted $P_C(v)$, to be the subgraph of $C$ that is formed by the union of all paths in $C$ terminating at $v$. Likewise, the \textbf{successor graph} at $v$, denoted $S_C(v)$, is the subgraph of $C$ that is formed by the union of all the paths in $C$ originating at $v$.
\end{definition}

Associated with any directed acyclic graph (DAG) $C = (V,E)$ is the
natural \textbf{vertex poset} $(V, \leq)$ where $u \leq v$ if and only if
there is a directed path from $u$ to $v$. Then the {\bf width of $C$},
denoted $\mathbf{w}(C)$, is the width of the poset $(V, \leq)$.

\begin{definition}
The \textbf{computation width} of a computational DAG $C = (V,E)$, denoted $cw(C)$, is defined as
$cw(C) = \max_{v\in V} \mathbf{w}(S(v))$.
\end{definition}

Let \OPT\ denote the optimal (off-line) algorithm. $W_\OPT(C)$ and $W_R(C)$ is the \emph{work} done by the optimal algorithm and a randomized algorithm $R$. We treat randomized algorithms as distributions over deterministic algorithms; for a set $\Omega$ and a family of deterministic algorithms $\{D_r \mid r \in \Omega \}$ we let $R = \mathcal{R}(\{ D_r \mid r \in \Omega \})$ denote the randomized algorithm where $r$ is selected uniformly at random from $\Omega$ and scheduling is done according to $D_r$. For a real-valued random variable $X$, we let $\Exp[X]$ denote its expected value. We let \OPT\ denote the optimal (off-line) algorithm. Specifically, for each $C$ we define $W_\OPT(C) = \min_D W_D(C)$.


\section{Lower bounds and Algorithm Modified-$RS$}\label{sec:lowerbound}

In this section we give a lower bound on our problem for $2$-level task graphs and we present the algorithm Modified-$RS$. We then show that for $2$-level task graphs the competitive ratio of Modified-$RS$ is tight. 

\begin{theorem}
\label{example}
Let $A$ be a scheduling algorithm for $2$-level task graphs, $\alpha$ be the fraction of tasks at level $l_1$. Then,  
\[
W_A \geq \left(1+ cw\left((1-\alpha) +  \frac{\alpha }{e^{\frac{1-\alpha}{\alpha} e+1}}\right)\right)W_{OPT} 
\] 
\end{theorem} 

\begin{proof}

Consider the 2 level task $t$-DAG where $G_1$  is the set of tasks at level $l_1$ and $G_2$ is the set tasks at level $l_2$ and the computation pattern described as follows . Initially, the computation pattern has $w$ groups  each consisting of a single processor. Let $t >> w$ and $t\mod w = 0$. Each processor completes $\frac{\alpha t}{w}$ tasks before they are merged into a single group $g(S)$ and allowed to exchange information about completed tasks before being split again into $w$ processors where each processor is allowed to complete $\frac{(1- \alpha) t}{w}$ tasks, at this point they are merged again into a single group $g(U)$ and then split into $w$ processors.
For this computation pattern the optimal off-line algorithm completes all the $t$ tasks at the formation of the group $g(U)$ and accrues exactly $t$ work.
Let $P_i \subset G_1$ denote the set of $\frac{\alpha t}{w}$ tasks for processor $i$. We analyze $A$ when the tuple $P = (P_1, \dots, P_w) $ is selected uniformly at random among all such tuples.
We will show that for any algorithm $A$ there is a configuration of the $P_i$ such that 
\[ W_A \geq \left(1+ (1-o(1))cw\left((1-\alpha) +  \frac{\alpha }{e^{\frac{1-\alpha}{\alpha} e+1}}\right)\right)t \]
Due to space restrictions we only give a sketch of the proof and we omit the details. We refer the reader to \cite{arxiv} for all the details. We first show that 
%
$\mathbb{E}[|L_S|] \geq \alpha t\left( 1 - \frac{1 }{w}\right)^w $
Where $L_S$ is the random variable whose value is the number of tasks of $G_1$ left undone at the formation of group $g(S)$.We then proceed by bounding the actual number of tasks left undone $T$ using Azuma's inequality and we show that $$\mathbb{E}[|L_U|] \geq (1-o(1))\frac{\alpha t}{e} \frac{1}{e^{\frac{1-\alpha}{\alpha} e(1 - o(1))}}$$ where $L_U$ is the random variable whose value is the number of tasks of $G_1$ left undone at the formation of group $g(U)$. In particular we show there must exist selection of the $P_i$ which achieves this bound. Note that after $g(U)$ the processors are split again into $w$ processors where they will complete the remaining \mbox{$(1-o(1))\frac{\alpha t}{e} \frac{1}{e^{\frac{1-\alpha}{\alpha} e(1 - o(1))}}$} tasks of $G_1$ and the $(1-\alpha)t$ tasks of $G_2$. This will give us the desired result.
\end{proof}

Note that when the tasks are independent ($\alpha = 1$) the lower bound is $1 + (1-o(1))\frac{cw}{e}$ which matches the result of  ~\cite{GRS_SIAM05} but the lower bound gets more pessimistic as the fraction of independent tasks gets smaller.

\subsection{Description and Analysis of Modified-$RS$}
%
%
%
In the following $l(t)  = i , i =1 \dots k$ denotes that task $t$ belongs to level $l_i$.  We are now ready to define Modified-$RS$ (m-$RS$) where a processor with knowledge that tasks in a set $K \subset V $ have been completed chooses the next task $\tau$ to be completed at random from $V \setminus K$ if and only if $\forall t \in V \setminus K$, $l(\tau)\leq l(t)$. In the following we analyze the competitive ratio of Modified-$RS$ and we show it's tight by obtaining the upper bound of the work performed by our algorithm on any computation pattern $(p,t)$-DAG and a $2$-level task $t$-DAG which matches the lower bound of the previous section.

\subsubsection{Upper Bound for m-$RS$ on a $2$-level task DAG}\label{sec:upperbound}

\begin{theorem}
\label{Theoremsplit}
Algorithm Modified-$RS$ is $\left(1+ cw\left(1-\alpha + \frac{\alpha}{e^{\frac{1-\alpha}{\alpha} c+1}}\right)\right)$-competitive for any computational $(p,t)$-DAG and for a 2-level task $t$-DAG. Here, $cw$ stands for the \textbf{computation width} of the computational $(p,t)$-DAG, $\alpha \in (0,1]$ ($\alpha$ is the fraction of tasks at level $l_1$) and $c = \frac{1}{\frac{1}{e} + o(1)}$.
\end{theorem} 

\begin{proof}
Due to space constraint we give an overview of the proof, we refer the reader to \cite{arxiv} for full details. We say a vertex $v$ in unsaturated if $\sum_{u<v}h(u) \geq t$, otherwise we say it is saturated. By linearity of expectation $ W_{m-RS} = \sum_{s \in \mathcal{S}}\mathbb{E} [T_s]  +  \sum_{u \in \mathcal{U}}\mathbb{E} [T_u]$, where $\mathcal{S} (\mathcal{U})$ is the set of saturated (unsaturated) vertices and where $T_v$ is the random variable denoting the number of tasks that m-$RS$ completes at vertex $v$. We construct the following bipartite graph $G = (\mathcal{S}, \mathcal{U}, E(G))$ s.t $E(G) = \{(s,u)| s< u \}$ and assign the weight $\mathbb{E}[T_v]$ to vertex $v$. We show that $\forall u \in \mathcal{U} \sum_{s\in \Gamma(u)}\mathbb{E}[T_s] = \sum_{s\in \Gamma(u)}h(s) \geq t$ and  $\forall u \in \mathcal{U} \sum_{u\in \Gamma(s)}\mathbb{E}[T_u] \leq cw\left(1-\alpha + \frac{\alpha}{e^{\frac{1-\alpha}{\alpha} c+1}}\right)t$. We then use a generalized degree-counting argument on the bipartite graph to show that $W_{m-RS} \leq \left(1 + cw\left(1-\alpha + \frac{\alpha}{e^{\frac{1-\alpha}{\alpha} c+1}}\right)\right) W_{OPT} $
To show the upperbound on $\sum_{u\in \Gamma(s)}\mathbb{E}[T_u]$ we use the following argument:  Consider $s \in \mathcal{S}$ a saturated vertex and it's successor graph $S(s)$. $S(s)$ is covered by $w$ paths $P_i, i = 1 \dots w$ where $w$ is at most $cw$. For each path $P_i$ let $u_i^0$ be the first unsaturated vertex and let $L_{u_i^0}$ be the random variable whose value is the set of tasks left incomplete by m-$RS$ at vertex $u_i^0$ we can show that $\mathbb{E}[|L_{u_i^0}|] \leq (1-\alpha)t + \frac{\alpha t}{e^{\frac{1-\alpha}{\alpha}c+1}}$ and thus $\sum_{u\in \Gamma(s)}\mathbb{E}[T_u] \leq cw( 1-\alpha+ \frac{\alpha }{e^{\frac{1-\alpha}{\alpha}c+1}})t$
\end{proof}

We extend the lower bound and upper bound results to any $k$-level task $t$-DAG. In theorem \ref{Theoremgen} we show the competitive ratio of our algorithm on any $k$-level task $t$-DAG, we also show that the ratio is tight. We refer the reader to \cite{arxiv} for detailed proofs.

\begin{theorem}
\label{Theoremgen}
Algorithm Modified-$RS$ is $\left(1+ cw\left((1-\alpha_1) +  \frac{\alpha_1 }{e^{\frac{\alpha_k}{\alpha_1} c^{a_k}+a_k}}\right)\right)$-competitive for any computational $(p,t)$-DAG and for any $k$-level task $t$-DAG where, $\alpha_i \in (0,1]$ and $c=\frac{1}{\frac{1}{e} + 1}$ and where $a_i, i=1 .. k$ is a sequence defined as follows, $a_1 = 1,  a_{i+1} = \frac{\alpha_i}{\alpha_1}c^{a_i} + a_i$
\end{theorem}

\section{Conclusions}
\label{sec:conc}

We studied the problem of cooperatively performing a set of $t$-tasks with dependencies in a decentralized setting where the communication medium is subject to dynamic changes. We pursued competitive analysis and presented a tight upper bound on the competitive ratio of our randomized algorithm Modified-$RS$ on $k$-level task $t$-DAG. When the tasks are independent our results subsume the results of ~\cite{GRS_SIAM05} and this bound is tight for the case of independent tasks. We show that the performance of any scheduling algorithm for leveled task graphs depends the computational width that captures the dynamics of the communication medium and on the nature of dependencies among the tasks.  In particular we show that performance of any algorithm in this model  can deteriorate as the size of the set of independent tasks reduces.



\end{document}